\documentstyle[prl,aps,epsf,multicol,amssymb]{revtex}
\begin{document}

%
\newcommand{\fig}[2]{\epsfxsize=#1\epsfbox{#2}}
%
%
%
\pagestyle{myheadings}
%
%
%
%
%
%
    \newcommand{\passage}{
   \end{multicols}\widetext\noindent\rule{8.8cm}{.1mm}\rule{.1mm}{.4cm}} 
   \newcommand{\retour}{
   \hspace{1cm}
   \noindent\rule{9.1cm}{0mm}\rule{.1mm}{.4cm}\rule[.4cm]{8.8cm}{.1mm}
   \begin{multicols}{2} }
   \newcommand{\unecol}{\end{multicols}}
   \newcommand{\deuxcol}{\begin{multicols}{2}}
%
%

\title{Glassy trapping of manifolds in nonpotential random flows}

\author{Pierre Le Doussal{$^1$} and Kay J\"org Wiese{$^2$}}
\address{{$^1$} CNRS-Laboratoire de Physique Th\'eorique de l'Ecole
Normale Sup\'erieure, 24 rue Lhomond 75231 Paris \cite{frad}}
\address{{$^2$} Fachbereich Physik, Universit\"at GH Essen, 45117 Essen, Germany}

\maketitle

\begin{abstract}
We study the dynamics of polymers and elastic manifolds
in non potential static random flows. We find that
barriers are generated from combined effects
of elasticity, disorder and thermal fluctuations. This leads
to glassy trapping even in pure barrier-free divergenceless
flows $v \stackrel{f \to 0}{\sim} f^\phi$ ($\phi > 1$). The physics is described by a
new RG fixed point at finite temperature.
We compute the anomalous roughness $R \sim L^\zeta$ and 
dynamical $t\sim L^z$ exponents for directed and isotropic manifolds.
\end{abstract}

\deuxcol


Elastic structures in presence of quenched 
disorder exhibit glassy behaviour 
\cite{vortex_review,kardar_review_lines,ledou_mglass}
such as pinning and very slow thermally activated 
motion of velocity $v \sim e^{-U(f)/T}$.
Understanding these properties is crucial for numerous
experimental systems, e.g. vortex lattices in
superconductors, magnetic domains,
solid friction and charge density waves
\cite{vortex_review,ledou_mglass}. 
For $D \le 4$ the static system is usually
in a disorder dominated glass state 
described by a $T=0$ RG fixed point \cite{fisher}. Barriers
diverge at small force $U(f) \sim f^{-\mu}$
and large scales \cite{vortex_review}. If $D \le 2$ thermal fluctuations
become important again. Periodic structures e.g. still exhibit a glass
phase in $D=2$, but with weaker logarithmic barriers
$v \sim f^\phi$ described by $T>0$ RG fixed points
\cite{cardy_desordre_rg,carpentier_desordre}.
Up to now, the study
of pinning and glassy dynamics has focused
on {\it potential} systems, typically
an elastic manifold in a random potential\cite{footnote1}.
The existence of pinning and barriers in a 
{\it non potential} system was proposed recently
in the context of vortex lattices driven over a
disordered substrate \cite{ledou_mglass}.
Driven systems even when all microscopic forces are formally
the gradient of a potential, are usually not so at a macroscopic
level since energy is constantly pumped into the system.
In \cite{ledou_mglass} it was found that
despite the non potential convective term present
in the equation of motion, some glassy properties 
survive. Related observations were made in (mostly mean field)
spin models \cite{cugliandolo_nonpotential_prl,ledou_rflow}.
Since little is known about general properties of
non potential systems with disorder, it is of great interest to study
other examples (and beyond mean field). An outstanding question is whether
the effective temperature, which tends to be generated 
in these systems from the constant dissipation, does overwhelm 
the effect of static disorder.

In this Letter, we study a model of a polymer in a 
non potential static random flow (a ``randomly driven
polymer'') and its generalization to an elastic manifold.
We work directly in finite dimension using dynamical RG.
We study two cases: either the polymer is directed
(each monomer sees a different flow) or it is isotropic
(all monomers are in the same flow) see Fig. \ref{fig1}. 
Our main finding is that the physics is described by
new RG fixed points with both $T>0$ and finite disorder,
leading to anomalous roughness $\zeta$ and glassy trapping
by the flow (with sublinear $v(f) \sim f^\phi$). Compared to 
the single particle studies \cite{ledou_diffusion_particle},
new features emerge such as the crucial role of
internal elasticity in generating barriers.
While being consistent with the
Hartree results ($d \to \infty$) for long range
(LR) disorder \cite{ledou_rflow}, our present
study yields new universal
fixed points for short range (SR) disorder.
It also generalizes the dynamics of self avoiding manifolds
\cite{kay_mope_dynamics} 
to a quenched disorder situation, though
self avoidance is perturbatively less
relevant in most regimes studied here.
For real polymers in realistic flows 
ours is mostly a toy model which
could be improved by including hydrodynamic
forces. Nevertheless, some of the physics unveiled 
here will be present in more realistic
models, such as the existence of preferred
regions in the flows. Inhomogeneities in polymer distribution
and non linearities in $v(f)$ could be
investigated experimentally.

\begin{figure}[htb]
\centerline{ \fig{8cm}{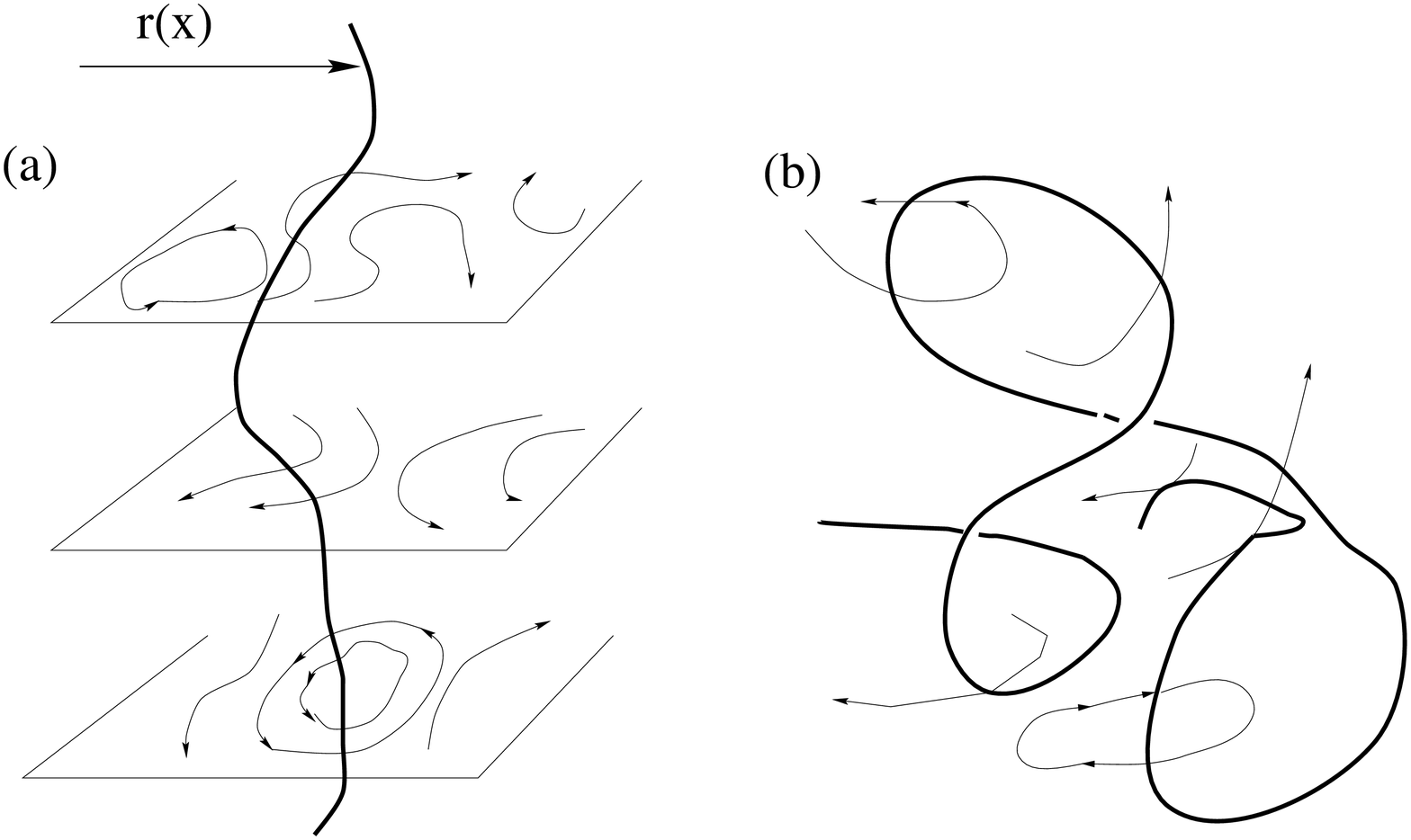} }
\caption{
{\narrowtext elastic manifolds (polymers $D=1$) in random flows:
(a) directed polymer (b) isotropic chain.
 \label{fig1} }}
\end{figure}

Let us illustrate how elasticity (or
entropy) leads to dynamical
generation of barriers in a divergenceless flow.
It is well known that for a single particle
convected in such a flow, the stationary measure 
at $T>0$ is spatially uniform and the drift velocity
under a force simply $v=f$. Remarkably 
as soon as one considers two elastically 
coupled particles, preferred regions appear.
Consider motion in the simple $2d$ flow ($v_x=-y$, $v_y=x$):
\begin{eqnarray}  \label{2particules}
&& \dot{z}_1 = c (z_2 - z_1) + i (\omega + \delta) z_1 + \eta_1  \nonumber \\
&& \dot{z}_2 = c (z_1 - z_2) + i (\omega - \delta) z_2 + \eta_2
\end{eqnarray}
$z_k = x_k + i y_k$ is the complex
position of particle $k$, $c$ the elastic coupling,
$<\eta_i \eta_j^*>=4 T \delta_{ij}$
is the thermal noise. Without disorder $\delta=0$ 
the motion is just a collective rotation
around the center (the matrix eigenvalues are $i \omega$
and $-2 c + i \omega$). At $T>0$ there is free collective diffusion
and $V=f$ as before. As soon as $\delta >0$
the zero mode disappears and the two particles converge
towards the center
(the eigenvalues are $\lambda_{\pm}=-c + i \omega \pm \sqrt{c^2-\delta^2}$).
This effect persists at $T>0$ since 
one finds $<|z_k|^2> = 2 T (c^2 + \delta^2)/(c \delta^2)$ ($k=1,2$) 
and thus there is a genuine bound state
despite the divergenceless nature of the flow. 
The stationary conformation is twisted since
$<z_1 z_2^*> = 2 T (c + i \delta)/\delta^2$. Extending 
(\ref{2particules}) to a directed chain, one
finds similar results (the decay towards the center
at $T=0$ is exactly the problem of spin depolarisation
\cite{mitra_ledou}). Thus a directed polymer in a flow will be
attracted to regions where 
elastic energy and dissipation are smaller (though the
precise balance remains to be understood). Presumably polymers
in realistic flows will be repulsed by high vorticity regions.
Non linear extensions of the above model
will thus show the generation of barriers. Similar effects
exist if all monomers see the same flow (isotropic chain).
This is reminiscent of the effect discovered 
by Fauve and Thual \cite{fauve}: mapping their 
Landau-Ginsburg function onto the complex position 
$z(x)$ of an elastic chain in a 2d non potential
non linear flow, one sees that (even at $T=0$)
it may converge to stable localized (rotating and twisted)
conformations $z(x,t) \to z_0(x)e^{i \omega t}$.

We now turn to our full disordered model where these
effects can be studied quantitatively. We consider a manifold
of internal dimension $D$ parametrized by a $d$-component field $r^{\alpha}(x)$.
The polymer corresponds to $D=1$
($x$ labels the monomers), and a single particle
to $D=0$. We study the Langevin dynamics:
\begin{equation} \label{depart}
\eta \partial_t  r^{\alpha}_{x t}  = c \nabla^2_x r^{\alpha}_{x t} + 
F_{\alpha}[r_{x t},x] + \zeta^{\alpha}_{x t}
\end{equation}
$\eta$ is the friction, $c$ the elastic 
coefficient and the gaussian thermal noise satisfies
$\langle \zeta^{\alpha}_{x t} \zeta^{\beta}_{x't'}  \rangle=
2 \eta T \delta_{\alpha \beta} \delta(t-t') \delta^D(x-x')$.
Angular brackets denote thermal averages and 
overbars disorder averages. 
$F_{\alpha}[r,x]$ is a gaussian quenched random force field
of correlations:
\begin{eqnarray}   \label{corr}
\overline{F_{\alpha}[r,x]  F_{\beta}[r',x']} 
= \Delta_{\alpha \beta}(r-r') h_{x-x'}
\end{eqnarray}
There are two main cases of interest. If the manifold
is {\it directed} (e.g. a polymer oriented by an
external field), then $h_{x-x'}=\delta^D(x-x')$. 
If the manifold is {\it isotropic}, (e.g. a gaussian chain in
a static flow), the force field does not depend
on the internal coordinate $F_{\alpha}[r,x] = F_{\alpha}[r]$
and $h_{x-x'}=1$. We consider a 
statistically rotationally invariant force field
with both a potential (L)
(``electric'') and a divergence-free (T)
part (``magnetic''). We thus study correlators 
$\Delta_{\alpha \beta}(r)=\int_K \Delta^{\alpha \beta}_K e^{iK.r}$
with, at small $K$:
\begin{equation}  \label{correlator}
 \Delta^{\alpha \beta}_K \sim K^{a-d} ( g_L \frac{K_{\alpha} K_{\beta}}{K^2}  + 
g_T (\delta_{\alpha \beta} - \frac{K_{\alpha} K_{\beta}}{K^2} ) )
\label{random force2}
\end{equation}
with $\int_K=\int \frac{d^dK}{(2 \pi)^d}$. 
We are mostly interested in
{\it SR correlated forces} $a=d$
(force correlator 
decaying faster than $r^{-d}$).
We also give results for {\it LR correlated forces}
(force correlator decaying as $r^{-a}$ with $a < d$). Note that
{\it SR random potentials}
$\Delta_{\alpha \beta}(K) \sim 
\tilde{g}_{\rm L} K_{\alpha} K_{\beta}
+ \tilde{g}_{\rm T} ( \delta_{\alpha \beta} K^2 - K_{\alpha} K_{\beta} )$
(formally $a>d$)
renormalize to the random force case, except 
in the potential case $g_T=0$, much studied
previously \cite{vortex_review,kardar_review_lines,fisher}
which is preserved by RG.

We study the time translational invariant
steady state. One defines the 
radius of gyration (or roughness) exponent
$\overline{\langle (r_{xt}-r_{x't})^2 \rangle}
\sim |x-x'|^{2 \zeta}$, the single monomer 
diffusion exponent $\overline{\langle (r_{xt}- r_{xt'})^2 \rangle}
\sim |t-t'|^{2 \nu}$ and assumes
scaling behaviour 
$\overline{\langle (r_{xt}-r_{00})^2 \rangle}
\sim |x|^{2 \zeta} \overline{B}(t/x^z)$
with $\zeta = z \nu$. The drift velocity under 
a small additional applied force $f$ in (\ref{depart}) is
$v \sim \overline{\langle r_{xt} \rangle}/t$.
We find $v \sim f^{\phi}$
at small $f$, with $\phi >1$ and thus the polymer
is trapped by the flow. Without disorder one has
$\zeta_0=(2-D)/2$, $\nu_0 = (2-D)/4$
for $D<2$, and bounded fluctuations for $D>2$.
\begin{figure}[htb]
\centerline{ \fig{7cm}{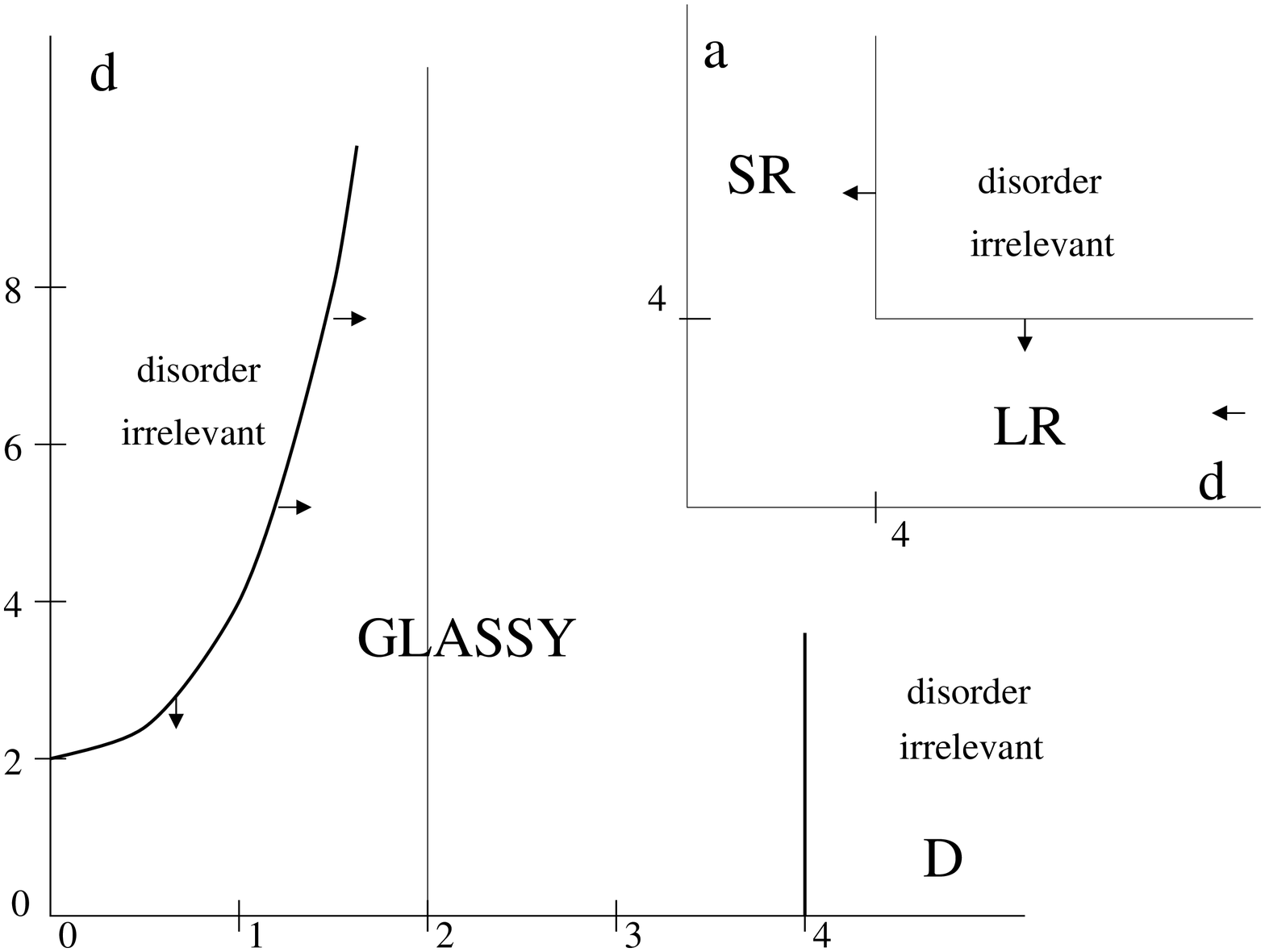} }
\caption{
\label{fig-correl}
{\narrowtext 
Regions in $d,D$ plane where SR disorder is relevant
for a directed manifold. Inset: Regions in $a,d$ plane 
where SR or LR disorder is relevant
for the directed polymer $D=1$.}}
\end{figure}
We start with simple Flory (i.e. dimensional) estimates. 
Balancing all terms in (\ref{depart}) (e.g. for directed manifolds
it reads $r/t \sim r/x^2 \sim r^{-d/2} x^{-D/2}$) yields:
\begin{equation} \label{flory}
\zeta^{\rm dir}_{\rm F} = 2 \nu^{\rm dir}_{\rm F} = \frac{4-D}{2+d}
~~~,~~~\zeta^{\rm iso}_{\rm F} = 2 \nu^{\rm iso}_{\rm F} = \frac{4}{2+d}
\end{equation}
for directed and isotropic manifolds, respectively
(with $d \to a$ for LR disorder). This suggests that the manifold
will be stretched by disorder \cite{footnote2}.
We now study (\ref{depart},\ref{corr}) using the dynamical generating functional
$Z= \int Dr D\hat{r} e^{- S_0[r,\hat{r}] - S_{{\rm int}}[r,\hat{r}] + i f \hat{r}}$
with:
\begin{eqnarray} \label{action2}
&& S_0 = \int d^Dx dt ~ i \hat{r}^{\alpha}_{xt} (\eta \partial_t - c \nabla_x^2)
r^{\alpha}_{xt}
- \eta T  i \hat{r}^{\alpha}_{xt} i \hat{r}^{\alpha}_{xt}  \\
&& S_{int}=
- \frac{1}{2} \int_{x,x',t,t'} (i \hat{r}^{\alpha}_{xt}) (i \hat{r}^{\beta}_{x't'})
\Delta_{\alpha \beta}(r_{xt} - r_{x't'}) h_{x-x'} \nonumber
\end{eqnarray}
We used dynamical RG methods,
both via a Wilson RG scheme (presented here) as in 
\cite{ledou_mglass,carpentier_desordre} and via a field theoretical method
using MOPE techniques as in \cite{sam_mope,kay_mope_dynamics}
(detailed in \cite{wiese_ledou_long}).
The free propagator is $B(x,t) = 
\frac{1}{2 D} \langle (r_{xt} - r_{00})^2 \rangle_0$ and the
free response function is 
$D R(x,t)=\langle i \hat{r}_{00} \cdot r_{xt} \rangle_0$.
(\ref{action2}, \ref{correlator}) is
invariant under the rescaling $x=e^l x'$, $t=e^{z l} t'$, $r=e^{\zeta l} r'$,
$\hat{r}=e^{(2 - z - \zeta - D + \beta)l} \hat{r'}$, provided
$\eta \to \eta e^{(2-z+\beta) l}$, 
$T \to T e^{(2-D-2 \zeta +\beta) l}$,
$c \to c e^{\beta l}$. Also 
$g_{T,L} \to g_{T,L}  e^{(4 - D + 2 \beta - (a+2) \zeta) l}$
(directed case) and 
$g_{T,L} \to g_{T,L}  e^{(4 + 2 \beta - (a+2) \zeta) l}$ 
(isotropic case) with $a=d$ everywhere for SR forces.
There are 
generically {\it three} independent exponents $z$, $\zeta$, $\beta$) and from 
them one gets:
\begin{eqnarray}  \label{phi}
\nu=\zeta/z ~~,~~ \phi = (z - \zeta)/(2- \zeta + \beta)
\end{eqnarray}
using that $f \to f e^{(D + z + \alpha) l}$, 
$v \to v e^{(\zeta-z) l}$ under rescaling.
Power counting at the Gaussian fixed point
with no disorder ($z=2$, $\beta=0$ and $\zeta=(2-D)/2$),
shows that disorder is relevant when (Fig.2)
$d<d_c^{\rm dir}$ for directed manifolds and 
$d<d_c^{\rm iso}$ for isotropic manifolds (resp. $a<d_c$ for LR
disorder), with:
\begin{eqnarray}
d^{\rm dir}_c = \frac{4}{2-D}~~~~ d^{\rm iso}_c = \frac{4+2 D}{2-D}
\end{eqnarray}
Thus for directed polymers disorder is relevant 
for $d<4$ and for isotropic polymers for $d<6$.
Power counting and symmetries show that the
only relevant terms generated in perturbation
theory are the one in (\ref{action2}).
To lowest order the
corrections to $T$, $c$ and $\eta$ read:
\begin{eqnarray} \nonumber
&& \delta T = \frac{(d-1)g_T}{\eta d}
\int_{\tau>0,y,K} h_y K^{a-d}
(e^{- K^2 B(y,\tau)} - e^{- K^2 B_{\infty}} ) \\
&& \delta c = \frac{g_L}{2 d D}
\int_{y,K} y^2 h_y K^{a-d}
(e^{- K^2 B(y,0)} - e^{- K^2 B_{\infty}} ) 
\label{temp}
\end{eqnarray}
and $(d-1) \delta \eta=(g_L/g_T) \eta \delta T$ (
where $B_{\infty}=\int_k \frac{T}{c k^2}$ is IR infinite
for $d \leq d_c$) \cite{footnote_calcul}. To simplify the above
expressions we have used that the free propagator satisfies
the FDT relation $\theta(t) \frac{d}{dt} B(x,t) = T R(x,t)$.
We note that a useful feature of the present model
is that although the FDT relation is not obeyed by the
exact Green's functions (since disorder is non potential)
it is within each scale of the RG. The FDT violation
will simply appear as a non trivial renormalization of $T$.
The relevant corrections to disorder read:
\begin{eqnarray}  \label{discorrections}
&&\delta \Delta^{\alpha \beta}_{P} =
\int_{x,y,\tau,\tau'}
\int_{K,K'=P-K}
( \Delta^{\alpha \beta}_K (K \cdot \Delta_{K'} \cdot K) \\
&& + (K' \cdot \Delta^{\alpha}_K) (K \cdot \Delta^{\beta}_{K'}) )
R_{x,\tau} R_{y,\tau'} h_{x-y} e^{K \cdot K' (B_{x,\tau} + B_{y,\tau'} )}
\nonumber
\end{eqnarray}
The small $P$ expansion in (\ref{discorrections}) is well behaved and because of
analyticity the only divergent terms
generated are proportional to $\delta_{\alpha \beta}$
yielding $\delta g_L = \delta g_T$.
One can thus take the limit $P \to 0$
($K'=-K$ in (\ref{discorrections})). One checks
explicitly from (\ref{discorrections}) that 
(i) starting with 
$\Delta_{\alpha \beta}(K) \sim 
\tilde{g}_{\rm L} K_{\alpha} K_{\beta}
+ \tilde{g}_{\rm T} ( \delta_{\alpha \beta} K^2 - K_{\alpha} K_{\beta} )$
with both $\tilde{g}_T>0$ and $\tilde{g}_L>0$ one generates SR random forces
and that (ii) a divergence free random force field
does not remain so.
The final RG equations have the form (specified to various cases
below and in Table 1):
\begin{eqnarray}   \label{rg}
&& \frac{d c}{c dl} = \beta + A_c \overline{g}_L ~~, ~~
\frac{d \eta}{\eta dl} = 2 - z + \beta + A_{\eta} \overline{g}_L \\
&& \frac{d T}{T dl} = 2 - D - 2 \zeta + \beta + A_T \overline{g}_T \\
&& \frac{d \overline{g}_L }{dl} =
\epsilon \overline{g}_L - (B-A) \overline{g}_L \overline{g}_T
+ E \overline{g}_L^2 \\
&& \frac{d \overline{g}_T }{dl} =
\epsilon \overline{g}_T + (A+E) \overline{g}_L \overline{g}_T
- B \overline{g}_T^2
\end{eqnarray}

\begin{figure}[htb]
\centerline{ \fig{6cm}{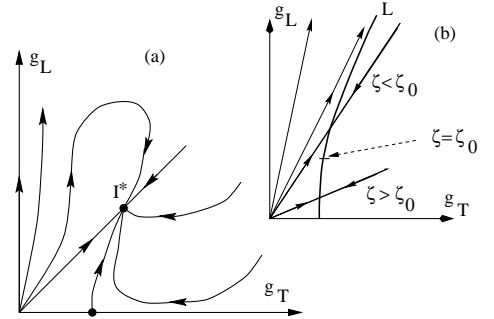} }
\caption{
\label{flowrg}
{\narrowtext
Examples of RG flow diagrams. (a) directed manifolds
with SR disorder. The physics is controlled
by the fixed point $I^*$ at $g_T=g_L$.
(b) isotropic manifolds LR disorder. The flow is
along straight lines. There is
a line of fixed points $L$ and an apparent
separatrix. $\zeta-\zeta_0$ changes sign
upon increasing $g_L/g_T$, suggesting progressive
localization. }}
\end{figure}

(i) {\it Directed manifolds, SR disorder}:
$\epsilon=2 - (2-D)\frac{d}{2}$.
Then $h_y=\delta^D(y)$ and there is no correction
to the elastic coefficient $\delta c=0$ (guaranteed
by the statistical tilt symmetry \cite{ledou_mglass}
for SR and LR disorder).
Thus $\beta=0$ and there are only two independent exponents.
The dimensionless
coupling \cite{footnote6} constant is 
$\overline{g}_{T,L} = C \Lambda^{\epsilon} g_{T,L}/T^{(d+2)/2}$.
The RG flow is depicted in Fig.3. There is a
globally attractive isotropic fixed point $I^*$ at 
$\overline{g}_L=\overline{g}_T=\epsilon/(B-A)$
(note that $B>A$) and
the line $\overline{g}_L=\overline{g}_T$ is preserved.
The potential line $g_T=0$
is preserved with a flow to strong coupling.
There is another {\it apparent} fixed point at $g_L=0, g_T = \epsilon/B$,
attractive along the line $g_L=0$ (divergenceless flows). However
for $D>0$ this line is not in the physical domain since,
as discussed above, a finite $g_L$ term is
generated and one flows to $I^*$. At $I^*$ one gets
\begin{eqnarray}  \label{result1}
&& z= 2 + \epsilon/a_{D^*} ~~,~~ 
\zeta = \frac{2-D}{2} + \epsilon/(2 b_{D^*}) \\
&& b_D =\frac{4-D}{2-D} - 
\frac{4}{D} \left( \frac{1}{2 \Gamma[D/2]} \right)^{2/(2-D)} ~,~
a_D = \frac{2+D}{2-D} b_D
\nonumber
\end{eqnarray}
and $\nu$ and $\phi$ using (\ref{phi}). The expansion can be carried
from any point of the line $d^*=4/(2-D^*)$.
Optimizing over $D^*$ as in \cite{sam_mope}
yields the estimates $\zeta = 0.625 \pm 0.02$,
$z=2.085 \pm 0.02$, $\phi=1.061 \pm 0.015$ (d=3) and 
$\zeta = 0.9 \pm 0.1$, $z=2.2 \pm 0.1$, $\phi=1.182 \pm 0.08$
(d=2). For small $\epsilon$, $\zeta>\zeta_F$ and
Flory is thus likely to be a lower bound
\cite{footnote_physics1}

(ii) {\it directed manifolds, LR disorder}: $\epsilon = 2 - \frac{(2-D) a}{2}$
with $a<d$.
The novelty is $\delta g_{L,R} = 0$ since
the LR part of the disorder is not renormalized
\cite{footnote4}. The dimensionless coupling \cite{footnote6}
$\overline{g}_{L,T} = C g_{L,T}/T^{(a+2)/2}$
will experience non trivial renormalization only
because $T$ renormalizes. Thus the ratio
$\frac{\overline{g}_L(l)}{\overline{g}_T(l)}= \frac{g_L}{g_T}$
is preserved which leads to a {\it line of fixed points}.
Thus $\overline{g}_T(l) \to g_T^{*}=\epsilon/B$ and 
$\overline{g}_L(l) \to \frac{g_L}{g_T} g_T^{*}$.
This yields the continuously varying exponent
$z = 2 + 2 \frac{g_L}{g_T} \frac{\epsilon}{(2 + a) (d-1)}$
and $\zeta=\zeta_F$ holds {\it to all orders}
in $\epsilon$ in the LR case since neither the
vertex nor $c$ renormalize. As conjectured in \cite{ledou_rflow}
$\nu$ is found identical to the Hartree approximation
(formula (12) \cite{footnote_hartree} of \cite{ledou_rflow}).

(iii) {\it isotropic manifolds (SR and LR disorder)}:
Then $\epsilon = D + 2 - a \frac{2-D}{2}$ ($a=d$ for SR disorder).
The novelty is a
renormalization of the elasticity of the manifold
(it becomes stiffer $\delta c >0$) and thus
a third non trivial exponent $\beta$.
The dimensionless constant is $\overline{g}_{T,L} = C
\frac{g_{T,L}}{T^{\frac{2+a}{2}} c^{\frac{2-a}{2}} }$.
For SR disorder the RG flow is similar to the directed case.
There is a fully attractive (isotropic disorder) fixed point
$I^*$ at $\overline{g}_T = \overline{g}_L = 1/(B-E-A)$.
As before, LR disorder is not renormalized,
the flow lines are $\frac{\overline{g}_L(l)}{\overline{g}_T(l)}=
\frac{g_L}{g_T} = k$ and there is a line of
fixed points $g_T = \epsilon /(B - k E)$, $g_L=k g_T$
parametrized by $k$.
The general formula for the exponent is
$\zeta=(2-D)/2 + \epsilon (A_T-k A_c)/2(B-kE-A)$,
$z=2 + k \epsilon (A_{\eta}- A_c)/(B-kE-A)$ and
$\beta=-k \epsilon A_c/(B-kE-A)$, with the values given 
in Table 1 (and set $k=1$ for SR). For SR disorder
this yields for $D=1$, $\zeta=0.5 + 0.13 \epsilon$,
$z=2 + 0.04 \epsilon$, $\beta=- 0.015 \epsilon$.
Extrapolations as in \cite{sam_mope} suggest
$\zeta= 0.85 \pm 0.1$, $z=2.1 \pm 0.1$, 
$\beta=-0.025 \pm 0.01$ for polymers in $d=3$.
Note that for LR disorder the line of fixed points is not
parallel to the axis any more (see Fig. 3), and thus there is {\it no fixed point}
for $k>k_c = B/E$. There is even a value $k^*=\frac{a-2}{a+2} k_c <k_c$ where 
$\zeta - \zeta_0$ changes sign. For $k^*<k<k_c$ 
the localization effects dominate. Together with
the apparent separatrix in the flow, this suggests
two phases (a localized one and another one with 
a continuously varying $\zeta$). It would be interesting
to check this scenario by non perturbative methods (such as in
\cite{ledou_rflow}).

Thus polymers (and manifolds) in static flows
are described by non trivial RG fixed points
at finite disorder. The main difference with conventional
static (potential) glasses is that a temperature is generated which
weakens disorder (and also renders the fixed point perturbatively
accessible - the coupling constant being
$g/T^{(d+2)/2}$ !). The main similarity lies in the sublinear $v(f)$
response reminiscent of the marginal glass
in the statics \cite{cardy_desordre_rg}. 
Numerical simulations would be helpful
to check the present results and further
investigate the physics of this
phase \cite{footnote_physics2}. The competition between
localization and driving found in some
cases for isotropic manifolds (i.e correlated disorder)
also deserves further studies.

We thank L. Cugliandolo, J. Kurchan, T. Giamarchi
and L. Sch\"afer for discussions.

\end{multicols}

$$
  \begin{array}{llllllllllllllllllll}
    &\vline\vline& 
A_{c} &\vline& 
A_{\eta} &\vline& 
A_{T} &\vline& 
B &\vline& 
E &\vline&
A  &\vline&
\cr \hline \hline
{\rm dir SR} &\vline \vline & 
0                     &\vline&
\frac{2}{d}           &\vline&
\frac{2(d-1)}{d}      &\vline&  
\frac{(2+d)}{2} A_T   &\vline&  
    0                 &\vline&  
\frac{16 \pi (d-1)}{d (d-2)(2-D) }
(\frac{S_D}{4 \pi})^{d/2}  &\vline&  
\cr \hline 
{\rm dir LR} &\vline \vline & 
 0                     &\vline&
\frac{2}{d}                 &\vline&
\frac{2(d-1)}{d}              &\vline&  
\frac{(2+a)}{2} A_T    &\vline&  
    0                  &\vline&  
     0                 &\vline& 
\cr \hline 
{\rm iso SR} &\vline \vline & 
\frac{1}{2 d D}             &\vline&
\frac{I_d}{d}               &\vline&
\frac{(d-1) I_d}{d}         &\vline&  
\frac{d+2}{2} A_T      &\vline&  
\frac{d-2}{2} A_c      &\vline&  
\frac{2 (d-1) \Gamma[\frac{D}{2-D}]^2}{
d (d-2)(2-D)^2 \Gamma[\frac{2D}{2-D}] } &\vline& 
\cr \hline 
{\rm iso LR} &\vline \vline & 
\frac{1}{2 d D}        &\vline&
\frac{I_a}{d}          &\vline&
\frac{(d-1) I_a}{d}    &\vline&  
\frac{a+2}{2} A_T      &\vline&  
\frac{a-2}{2} A_c      &\vline&  
0                      &\vline& 
\end{array} 
$$
{\small 
{\bf Table I}: Coefficients of the RG equation
(\ref{rg})
for cases studied here (one must set
$d=d^*$ and $D=D^*$). One denotes $I_a = \int_0^{+\infty} dt
\tilde{B}(1,t)^{-a/2}$ with
$\tilde{B}(1,t) = \frac{1}{\Gamma[D/2]}
((4 t)^{(2-D)/2} e^{-1/(4t)} + \Gamma[D/2,1/(4t)])$, with
$\tilde{I}_{d=6} \approx 1.7935$,
\widetext
\begin{multicols}{2}


\unecol

\begin{thebibliography}{999}
\bibitem{footnote1} or hamiltonian systems (which have similar
properties).
\bibitem{footnote2} however $\zeta = \zeta_{\rm F}$ is
exact {\it only} for directed manifolds with LR correlations
with also $\nu = \nu_{\rm F}$
$g_L=0$ and $d \to \infty$ (while
for $D=0$ and $g_L=0$, $\nu = \nu_{\rm F}$ is {\it always} exact
\cite{ledou_rflow}).
\bibitem{footnote4} (\ref{discorrections}) renormalizes only the SR part of disorder
and becomes important only in the region of the $(a,d)$ plane 
where the crossover\cite{ledou_diffusion_particle} LR-SR occurs (not studied here).
\bibitem{footnote5} 
We have extended the calculation
of \cite{ledou_rflow} and obtained the $\phi$ exponent for Hartree
in agreement with (\ref{phi}).
\bibitem{footnote6} We use
$C=\frac{\Gamma[a/2]}{\Gamma[d/2]}
(4 \pi)^{\frac{a-d}{2} -1} (\frac{2-D}{2})^{a/2}$
and $C'= (2-D)^{a/2} \frac{\Gamma[a/2]}{\Gamma[d/2]} S_D^{1+\frac{a}{2}} 
(4 \pi)^{-d/2}$ with
$a=d$ for SR disorder.
\bibitem{footnote_calcul}
(\ref{temp},\ref{discorrections}) are computed using 
$B(x,0) = T |x|^{2 - D}/((2-D)S_D c)$,
$B(0,t) = 2 T |\mu t|^\frac{2 - D}{2}/((2-D)(4 \pi c)^{D/2})$
with $S_D = 2 \pi^{D/2}/\Gamma(D/2)$ and
$\int_K e^{- K^2 B} K^{a-d} =$ 
$( 4 \pi )^{-d/2} \frac{\Gamma[a/2]}{ \Gamma[d/2]} B^{-a/2}$.
\bibitem{particle}
Continuing (\ref{result1})
to $D=0$ yields $\zeta=1 + \frac{\epsilon}{2}$,
$z = 2 + \epsilon$ (with $\epsilon=2-d$).
This yields $\nu=\zeta/z=1/2 + O(\epsilon^2)$
and $\phi=1 + \epsilon$ at $I^*$ which is the
correct known result for the particle, obtained via 
a different RG \cite{ledou_diffusion_particle}.
\bibitem{footnote_physics2}
A scenario to be checked is whether the hierarchy
of wells within wells of a static glass is interrupted here
at some random level by the nonpotential forces (a mechanism
to generate wells with algebraic distribution of
release time) and if there are Levy walk type of jumps
between trapped configurations.
\end{thebibliography}
\end{document}